\documentclass[11pt,twocolumn]{scrartcl}
\usepackage{casa_conf}	% The CASA style
\usepackage{graphicx}	% A package for graphics use (see figures)

\usepackage{amsopn}
\usepackage{amsmath}
\usepackage{amsfonts,amssymb} 
\usepackage{siunitx}
\usepackage{subfigure}

\usepackage{makecell}
\usepackage{multirow}
\usepackage{xcolor}
\usepackage{mathrsfs}
\usepackage{epstopdf}
\usepackage[switch]{lineno}

\title{FASTSWARM: A Data-driven \textbf{F}r\textbf{A}mework for Real-time Flying In\textbf{S}ec\textbf{T} \textbf{SWARM} Simulation}

\author{Wei Xiang\textsuperscript{1} \and Xinran Yao\textsuperscript{1} \and He Wang\textsuperscript{2} \and Xiaogang Jin\textsuperscript{1,3}\thanks{Corresponding author. E-mail: jin@cad.zju.edu.cn}\\
        \and
        \normalsize \textsuperscript{1} State Key Lab of CAD\&CG, Zhejiang University, Hangzhou 310058, China\\
        \and
        \normalsize \textsuperscript{2} School of Computing, University of Leeds, Leeds LS2 9JT, United Kingdom\\
        \and
        \normalsize \textsuperscript{3} ZJU-Tencent Game and Intelligent Graphics Innovation Technology Joint Lab, Hangzhou 310058, China
      }
\begin{document}
\maketitle
%-------------------------------------------------------------------------
\begin{abstract}
Insect swarms are common phenomena in nature and therefore have been actively pursued in computer animation. Realistic insect swarm simulation is difficult due to two challenges: high-fidelity behaviors and large scales, which make the simulation practice subject to laborious manual work and excessive trial-and-error processes. To address both challenges, we present a novel data-driven framework, FASTSWARM, to model complex behaviors of flying insects based on real-world data and simulate plausible animations of flying insect swarms. FASTSWARM has a linear time complexity and achieves real-time performance for large swarms. The high-fidelity behavior model of FASTSWARM explicitly takes into consideration the most common behaviors of flying insects, including the interactions among insects such as repulsion and attraction, the self-propelled behaviors such as target following and obstacle avoidance, and other characteristics such as the random movements. To achieve scalability, an energy minimization problem is formed with different behaviors modelled as energy terms, where the minimizer is the desired behavior. The minimizer is computed from the real-world data, which ensures the plausibility of the simulation results.
Extensive simulation results and evaluations show that FASTSWARM is \textit{versatile} in simulating various swarm behaviors, \textit{high fidelity} measured by various metrics, easily \textit{controllable} in inducing user controls and highly \textit{scalable}.
% We present a data-driven optimization method to simulate plausible animations of flying insect swarms in real time. To generate the trajectories of flying insects, we use an interactive optimization-based method to update the motion state of each agent by employing a motion characteristic dataset as the solution space. To generate the reference dataset, we take the velocity and the magnitude of acceleration of an insect for each time step in the input motion as the motion characteristics. Our method then updates the motion state for each agent by selecting a data term minimizing an optimization function from the reference data. The terms in the optimization function are designed according to numerical analyses and empirical observations about flying insects, including the interaction among agents, the self-propulsion of agents, the noise-induced control, and the user-defined control. Our approach provides a general framework for simulating swarms of flying insects with plausible motion, and is scalable to any number of flying insects in various environments. Experimental results show that our method is able to simulate various insect swarm behaviors including aggregation, mating, escaping, and trajectory constraints by setting different parameters. In addition, our method can be used to predict the missing motion data in the captured trajectories of insect swarms or generate augmented flying insect swarms with a different number of insects as in the input.  
\end{abstract}
\linebreak
\linebreak
\section{Introduction}
\label{sec:introduction}
Insects are ubiquitous in both the real and virtual worlds, and many of them present collective behaviors for efficient and collaborative work. In the real world, flying insect swarms can exhibit a great variety of behaviors such as aggregation, mating, migration and escaping \cite{sinhuber2017phase}, where the individual behaviors are often correlated in various ways, from collaborative to competitive or even adversarial. Simulating realistic insect swarms are in the interest of many areas. In robotics, research of insect swarms has led to new algorithms for robots’ collective jobs on information transfer, decision-making, task partitioning or transport \cite{blum2008swarm,halloy2007social}. In computer animation, insect swarms have been used to create wondrous natural phenomena and interesting visual effects \cite{wang2014inherent,chen2019shape}. However, simulating scalable collective behaviors of insect swarms with high-fidelity remains challenging.

Existing methods for simulating flying insect swarms mainly fall into two categories: empirical and data-driven. Empirical methods aim to abstract swarm behaviors into mathematical models and deterministic systems, such as the field-based methods \cite{wang2014inherent}, or a combination of the field-based with the force-based methods \cite{chen2019shape}. One limitation of such methods is that the simulated trajectories are often too regular and lack of visual diversity, due to their determinism nature. In contrast, data-driven methods tend to rely on real-world data, such as using computer vision techniques to capture 3D trajectories of swarms \cite{kelley2013emergent,puckett2014searching,wu2011automated} for simulation \cite{li2015biologically,wang2015bswarm,ren2016simulating}. However, due to the intrinsic limitations of optical sensors (e.g. occlusions), the motion capture is set up in massively simplified laboratory environments, and there are still excessive tracking errors where only short tracklets can be relatively reliably obtained. This creates tremendous difficulties in simulating flying insect swarms with the desired high-fidelity and scalability. First, the captured trajectories cannot be relied upon to extract all behaviors of flying insects. Second, the generalizability of the model based on simple data is limited by both the environment complexity and the swarm size.

In this paper, we propose a novel data-driven framework (FASTSWARM) to address the challenges for simulating flying insect swarms. Our framework models insects as agents, and the swarm behavior computation as an energy minimization problem. A variety of important behaviors identified in numerical analysis \cite{Lukeman12576} and empirical observations are captured by different energy terms, including the interaction among agents, the self-propulsion of agents and the motion noise of agents, so that the minimizer leads to realistic behaviors.
Besides, our framework also model user-defined behaviors by employing user-control energy terms. The total energy function is constructed in the way that it can be optimized quickly to achieve scalability and real-time performance. 

During optimization, instead of seeking a minimizer by pure mathematical optimizations which would make the minimizer only ideal in theory, we seek the minimizer by referencing a motion characteristic dataset generated from the real-world data, so that the simulated behaviors mimic the real data. However, this means that the motion characteristics we reply on in the reference dataset has to be reliable. In the real-world data, although excessive noises exist and whole trajectories can rarely be obtained, velocity is much more reliable as it can be estimated from short tracklets \cite{ren2019heter}. In our simulation framework, both velocity and acceleration are regarded as the motion characteristics for generating the reference dataset, and we therefore optimize for the velocity to update the motion states of the agents. In addition, we use an implicit Euler scheme to improve the stability.

Formally, the contributions of the paper include:
\begin{itemize}
	\item A novel data-driven 3D swarm simulation framework which captures a variety of biologically important behaviors.
	\item An optimization method that maximally makes use of real-world data to ensure the simulation fidelity for flying insect swarms.
	\item A scalable model for large swarm simulation with straightforward user control.
\end{itemize}

The remainder of this paper is organized as follows. After briefly introducing related work in Section \ref{sec:related work}, we give a pipeline overview of our approach in Section \ref{sec:overview}. In Section \ref{sec:method}, we explain our optimization-based data-driven model. We show simulation results and evaluations of our method in Section \ref{sec:results and evaluations}, and discuss the limitations and future work in Section \ref{sec:conclusion}.
\begin{figure*}[t]
	\centering
	\includegraphics[width=\linewidth]{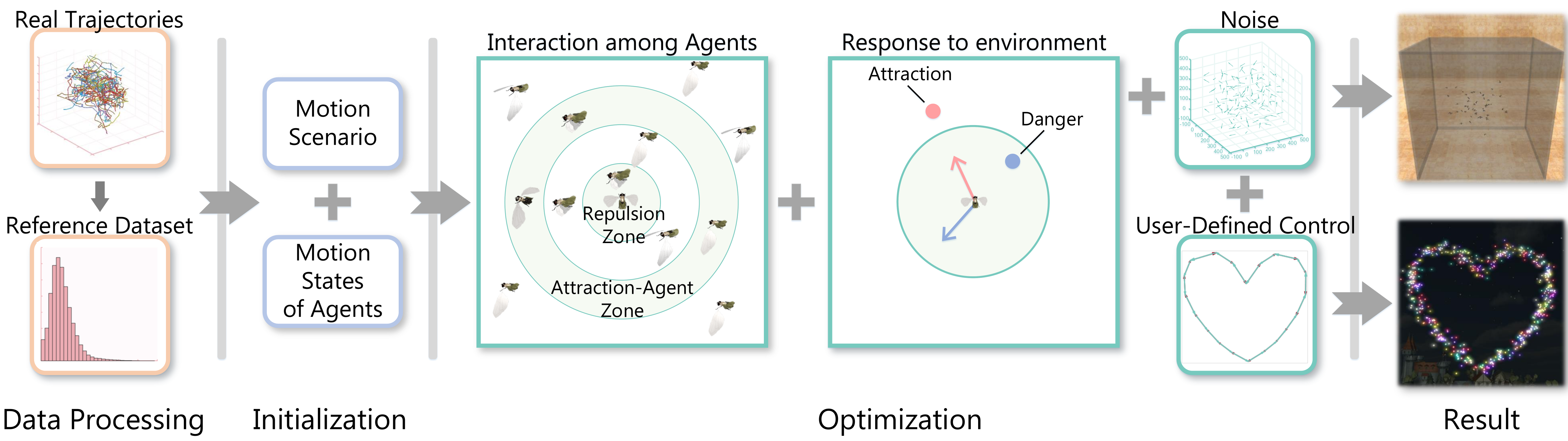}
	\caption{\label{fig:framework}
		Overview of our data-driven approach for simulating flying insect swarms.}
\end{figure*}
\section{Related Work}
\label{sec:related work}

\textbf{Data-Driven Simulation}. In graphics, data-driven methods have been proposed to simulate behaviors of crowds and traffics. Given trajectories (or tracklets) extracted from crowd data, example-based methods can blend them to generate new animations \cite{ju2010morphable}, use a ``clone and paste'' technique to generate larger crowds \cite{li2012cloning}, or cluster them into groups and update the motion of an agent based on the actions of its nearest patch or associated group \cite{lee2007group,zhao2013data,charalambous2014pag}. 

In data-driven traffic simulation, Chao et al. \cite{chao2013video} present a video-based approach to learn the specific driving characteristics of drivers to reconstruct or simulate traffic flows. By taking the spatio-temporal information of traffic flows as a 2D texture, a texture synthesis technique is developed to populate virtual road networks with realistic traffic flows  \cite{chao2017realistic}. In addition, deep learning can also be used to learn the latent patterns of vehicle trajectories for intersectional traffic simulation and editing \cite{bi2019deep}. Recently, an interactive data-driven optimization approach \cite{ren2019heter} has been proposed to simulate traffic scenarios with heterogeneous agents. 

Although these methods can generate plausible crowd or traffic animations, they focus on 2D simulation and cannot be easily extended to 3D flying insect swarms because the motion dynamics of insects are significantly different.

\textbf{Insect Swarm Simulation}.
There has been an interdisciplinary effort in the research of collective behaviors of insects. Researchers in agriculture proposed an insect migration trajectory simulation method to accurately predict the destinations of insect migration and achieve effective early warning to reduce the impact of pests on agriculture \cite{wang2019insect}. Public health researchers proposed an indoor flight behavior model of host-seeking mosquitoes for selecting bed nets that can effectively reduce the spread of the virus from mosquitoes \cite{jones2020minimal}. In swarm robotics, swarm behaviors can be used to control collaborative robots, such as the insect-based biobots for search and rescue \cite{bozkurt2014biobotic}, unmanned aerial vehicle quadrotors emulating insect swarm behaviors \cite{bandala2014swarming},
and the task allocation algorithm based on ant colonies \cite{khaluf2019local}.

In graphics, methods have been proposed to simulate collective behaviors of ants \cite{guo2017simulating,xiang2019biologically}. However, it is non-trivial to extend these methods to simulate flying insect swarms because of their significant differences in motion dynamics. Data-driven methods have been developed to simulate flying insect swarms. Li et al. \cite{li2015biologically} present a framework to simulate flying insect swarms by sampling the statistical information from real datasets. However, it is not general enough because a new local steering model is needed whenever the data changes. Different from the prior work, we present a general framework which is maximally data-driven and automatically adaptive to data, and therefore is not tied to specific data. Recently, data-driven noise models and force-based models are introduced in \cite{wang2015bswarm,ren2016simulating} to generate biologically plausible animations of flying insect swarms. However, they do not generalize well to more complex scenarios because they are prone to numerical errors during generalization.

Except for data-driven methods, a hybrid model combining potential fields and curl-noise \cite{bridson2007curl} is developed in \cite{wang2014inherent} to simulate various behaviors of flying insect swarms. However, such a forced-based and field-based method often generates trajectories that look too regular because all agents share similar motion patterns. For the special effect simulation application, Chen et al. present a flock morphing method of flying insect swarms with pre-defined shape constraints \cite{chen2019shape}. Instead of pre-defined morphing, our framework can generate plausible animations based on real-world motions to satisfy user-defined constraints.

\section{Methodology Overview}
\label{sec:overview}
%Our method calculates trajectories for the agents to generate various insect swarm animations by referencing datasets to solve an optimization problem, Figure \ref{fig:framework} illustrates the
%process of our simulation approach.
As illustrated in Figure \ref{fig:framework}, FASTSWARM can be conceptually described as a three-phase process: the data processing stage, the initialization stage, and the real-time simulation stage.
Our simulation algorithm makes full use of the dataset extracted from the data processing stage to compute the trajectories of the agents in the swarm. At the data processing stage, the raw data are noisy tracklets
\cite{kelley2013emergent,wu2011automated}.
For each type of insects, we generate a reference dataset that contains the motion characteristics specific to that type. Each data sample consists of two items: velocity and acceleration, estimated using forward differencing. We organize the data by the speed, similar to \cite{ren2019heter}, for fast indexing. At the initialization stage, we initialize the motion scenario, the number of insects, and the initial motion state of each agent. The motion state of an agent includes its position, velocity, motion randomness, and control direction. During simulation, we employ a data-driven approach to update the motion state for each agent. For every time step, our model selects
a velocity from the reference dataset that minimizes an objective function which models the interactions, self-propulsion, and user-specified controls. 
% In addition, the objective function is designed based on the numerical analysis \cite{Lukeman12576} and empirical observations of insects, including the interactions, the self-propulsion, and the noise-induced motion control. 

% Specifically, the interaction among agents includes the repulsion and the attraction among agents, the self-propulsion of agents includes the internal drive and the response to environment, and the noise-induced motion control introduces random disturbance to the moving direction of each agent.
% To enrich the diversity of the simulation results, a user-defined control term is also introduced to enforce additional constraints to the agents.

%We use the empricial data from \cite{kelley2013emergent} that was captured in a transparent cubical enclosure with laboratory swarms of the midge \emph{Chironomus riparius}.
%The dataset contains the trajectories of all midges in the manner of different sequential position information.

%Then we initialize the position of each agent in the scene randomly and choose an initial velocity for each agent from our datasets.

%At at each timestep of the real-time simulation stage, we employ a data-driven algorithm to update the motion

\section{Data-Driven Optimization Model}
\label{sec:method}
The reference data is denoted as $D = \cup_{v} d_v$, where
$d_v = \left[\mathbf{v}, \mathbf{a} \right]$,  $\mathbf{v}$ is the velocity, and $\mathbf{a}$ is the acceleration.
For a swarm with $N$ agents at time $t$, the state of agent $i\ (i=1,...,N)$ is denoted as
$\mathbf{s}_{i, t} = [\mathbf{p}_{i, t}, \mathbf{v}_{i, t}, \hat{\mathbf{v}}^n_{i, t}, \hat{\mathbf{v}}^{cd}_{i, t}]$, $\mathbf{s}_{i,t} \in \mathbb{R}^{12}$, where $\mathbf{p}_{i, t} \in \mathbb{R}^{3}$ is the current position,
$\mathbf{v}_{i, t} \in \mathbb{R}^{3}$ is the current velocity, $\hat{\mathbf{v}}^n_{i, t} \in \mathbb{R}^{3}$ is the noise direction, and $\hat{\mathbf{v}}^{cd}_{i, t}$ is the user-defined control direction. 
%To be specific, the noise induces randomness to the agent's moving direction.
We further use $\mathbf{S}_{t} = \cup_{i}{s}_{i, t}$ to represent the motion state of the whole swarm at $t$. Then, the motion dynamics of agent $i$ is formulated as:
\begin{equation}
\label{equ:updating function}
\begin{array}{l}
\mathbf{v}_{i, t+1} = \mathop{\arg\min}\limits_{\mathbf{v}_{i, t+1} \in {d_v} \in D}{E(i, \mathbf{v}, \mathbf{a}, \mathbf{S}_{t})}, \\
\mathbf{p}_{i, t+1} = \mathbf{p}_{i, t} + \mathbf{v}_{i, t+1} \Delta t, \\
\hat{\mathbf{v}}^n_{i, t+1} = f_{N}(\mathbf{s}_{i, t}), \\
\hat{\mathbf{v}}^{cd}_{i, t+1} = f_{CD}(\mathbf{s}_{i, t}, ENV_{t}), \\
\end{array}
\end{equation}
where the velocity $\mathbf{v}_{i, t+1} \in d_v \in D$ minimizes the objective function $E$, $\mathbf{a}$ is the acceleration in $d_v$, $\Delta t$ is a timestep, and $\mathbf{p}_{i, t+1}$ is the position at $t+1$.
After updating the velocity and position of each agent, we update the noise direction $\hat{\mathbf{v}}^n_{i, t+1}$ and the user-defined control direction $\hat{\mathbf{v}}^{cd}_{i, t+1}$ using $f_{N}$ and $f_{CD}$, respectively. The objective function $E$ is defined as:
\begin{equation}
\label{equ:E}
E(i, \mathbf{v}, \mathbf{a}, \mathbf{S}_{t}) = E_{i, int} + E_{i, sp} + E_{i, n} + E_{i, user},
\end{equation}
where $E_{i, int}$ is an interaction term to model the interactions among agents, $E_{i, sp}$ is an self-propulsion term to control the motion of agents,  $E_{i, n}$ is an noise-induced control term to model the randomness of agents' movements, and $E_{i, user}$  is the user-control term to enforce user-defined constraints.
%and $w_{n} \in \mathbb{R}$, $w_{user} \in \mathbb{R}$ are the weights of $E_{i, n}$ and $E_{i, user}$ respectively .

For each agent $i$, the minimization of $E$ aims to update $\mathbf{v}_{i, t+1}$ by selecting a $\mathbf{v} \in d_v \in D$. For simplicity, we use the selected velocity $\mathbf{v}$ and $\mathbf{\hat v}$ to represent $\mathbf{v}_{i,t+1}$ and its direction in the following definitions of the energy terms.

\subsection{Interaction among Agents}
Insects move in coordination with each other and aggregate without collisions. In FASTSWARM, such coordination is modeled as interactions which include repulsion for collision avoidance and attraction for aggregation. The interaction energy $E_{i, int}$ is therefore defined as:
\begin{equation}
\label{equ:interaction}
E_{i, int} = E_{i, rep} + E_{i, att},
\end{equation}
where $E_{i, rep}$ and $E_{i, att}$ represent the repulsion energy and the attraction energy, respectively. 
%where $E_{i, rep}$ is the repulsion term and $E_{i, att}$ is the attraction term, $w_r \in \mathbb{R}$ and $w_a \in \mathbb{R}$ are the weights of the two terms respectively.

%\begin{figure}[ht]
%	\centering
%	\subfigure[Interaction range for agents]{
%		\label{subfig:interaction_range} 
%		\includegraphics[width=0.45\linewidth]{figures/method/interaction_range.pdf}}
%	%\hspace{0.05in}
%	\subfigure[Response range for agents]{
%		\label{subfig:response_range} 
%		\includegraphics[width=0.45\linewidth]{figures/method/response_range.pdf}}
%	\caption{\label{fig:ranges} The circular zones for an agent. In both Figure \ref{subfig:interaction_range} and Figure \ref{subfig:response_range}, the 
%		the 
%		Based on the relative distance for the agents with their neighbors, the repulsion happens when the distance is in a short range $d_{re}$,
%		and the attraction happens when the distance is far from a boundary distance $d_{att}$, and $d_{att} > d_{rep}$. 
%	}
%\end{figure}

\subsubsection{Repulsion}
An agent avoids collisions with other agents that are too close to it. Based on the zonal interaction model, which dictates that an insect has a short-range repulsion within a circular distance-based zone \cite{Lukeman12576}, we describe the \textit{repulsion zone} using a spherical kernel function, centered at the agent with a predefined cut-off radius $d_{rep}$ (see Figure \ref{fig:framework} for the 2D representation), and an agent will avoid collisions with its neighboring agents within the \textit{repulsion zone}.
%For agent $i$,
Assuming that agent $i$ will move in a timestep using velocity $\mathbf{v}$ in the data term $d_v$ and it's repulsive neighbor $j$ will hold its current velocity $\mathbf{v}_{j, t}$ in a timestep, then the distance-based repulsion energy $E_{i, rep}$ is calculated as follows:
\begin{equation}
\label{equ:repulsion}
\centering
\begin{array}{lr}
\mathbf{p}'_{i, t+1} = \mathbf{p}_{i, t} + \mathbf{v} \Delta t, \\
\mathbf{p}'_{j, t+1} = \mathbf{p}_{j, t} + \mathbf{v}_{j, t} \Delta t, \\
\begin{split}
E_{i, rep} =  &w_{rep} \cdot \\
&\frac{1}{|RN|} \sum\limits_{j \in RN} e^{\rho \left(1 - \frac{\|\mathbf{p}'_{i, t+1} -\mathbf{p}'_{j, t+1}\|_2}{d_{rep}} \right)},
\end{split}
\\
\end{array}
\end{equation}
where $w_{rep} \geq 0$ is the weight of the repulsion energy, $RN$ is the set of neighbor agents, $|RN|$ is the number of the repulsive neighbors, $\rho > 0$ is a constant used for the scaling of the energy, and $\mathbf{p}'_{i, t+1}$ and $\mathbf{p}'_{j, t+1}$ are the predicted positions of agent $i$ and its neighbor $j$. 

% predicts the position of agent $i$ using the chosen velocity $\mathbf{v}$ after a timestep, and  is the predicted position of the repulsive neighbor $j$ under the assumption that $j$ will hold its velocity in a timestep. The repulsion energy is normalized by averaging the sum of the energies generated by the neighbors.

%The repulsion energy $E_{i, rep}$ decreases as the predicted distance in Equation \ref{equ:repulsion} gets larger.

\subsubsection{Attraction}
Attraction exists between agents, and also between every agent and the center of the swarm  \cite{Lukeman12576}. The attraction energy $E_{i, att}$ therefore consists of both the attraction from the neighbors and the attraction from the swarm center:
\begin{equation}
\label{equ:attraction}
\begin{array}{l}
E_{i, att} = E_{i, oa} + E_{i, cos}, \\
\end{array}
\end{equation}
where $E_{i, oa}$ is the energy of attraction from the neighbors, and $E_{i, cos}$ is the attraction energy from the center of the swarm. As shown in Figure \ref{fig:framework}, the attraction zone of an agent to its neighbors is also formulated as a spherical kernel function with two cut-off radii, one inner $d_{att, 1}$ and one outer $d_{att, 2}$:
%and the attraction
% an insect is attracted by the other insects whose distance to the it is between the range $d_{att}$ to $d^\prime_{att}$.
The attraction energy from the other agents $E_{i, oa}$ is defined as:
\begin{equation}
\label{equ:attractAgent}
E_{i, oa} = w_{oa} \cdot \frac{1}{|AN|}\sum\limits_{j \in AN}e^{\rho \left(\frac{\|\mathbf{p}'_{i, t+1} - \mathbf{p}_{j, t}\|_2}{d_{att,1}} - 1 \right)},
\end{equation}
where $w_{oa} \geq 0$ is the weight of the energy term, $AN$ is the set of the attraction neighbors of agent $i$, and $|AN|$ is the number of attraction neighbors.
The attraction energy on agent $i$ is normalized by the total number of the neighbor agents.
%
%% $E_{i, oa}$ gets larger as the predicted distance between the agent's predicted position $\mathbf{p}'_{i, t+1}$ to the attraction neighbor $\mathbf{p}_{j, t}$ gets larger.

For the attraction to the swarm center on agent $i$, we use $\hat{\mathbf{v}}_{i, center}$, the direction that points from the agent's current position to the swarm center, as the desired direction of the chosen velocity $\hat{\mathbf{v}}$.
The attraction from swarm center $E_i^{cos}$ is defined as:
\begin{equation}
	\label{equ:attractCenter}
	E_{i, cos} = w_{cos} \cdot e^{\rho \| \hat{\mathbf{v}} - \hat{\mathbf{v}}_{i, center} \|_2},
\end{equation}
where $w_{cos} \geq 0$ is the weight of the energy term.
%Analogously, the energy of attraction from the center of the swarm is calculated as
%\begin{equation}
%\label{equ:attractC}
%E_{i, cos} = e^{\rho \left( \frac{\|\mathbf{p}'_{i, t+1} - \mathbf{p}_{cos, t}\|_2}{d_{cos}} - 1\right)},
%\end{equation}
%where $\mathbf{p}'_{i, t+1}$ is the focal agent's predicted position in Equation \ref{equ:repulsion} and $\mathbf{p}_{cos, t}$ is the centroid position of the swarm at time $t$.
%
%The attraction energy term $E_{i, cos}$ gets larger as the predicted distance in Equation \ref{equ:attractC} gets larger.
\subsection{Self-Propulsion}
Besides being reactive, insects are also self-propelled. We therefore define a self-propulsion energy term, not only to drive the agents to keep moving similar to the reference trajectories, but also to response to external stimuli in environments. For agent $i$, the self-propulsion energy includes the internal propulsion term $E_{i, vel}$ to drive the agent to generate new trajectories similar to the data, and the reaction term $E_{i, env}$ to drive the agent to react to external stimuli:
\begin{equation}
E_{i, sp} = E_{i, vel} + E_{i, env}.
\end{equation}
%where $w_{vel} \in \mathbb{R}$ and $w_{env} \in \mathbb{R}$ are the weights of the two energy terms respectively.
%where $E_i^{vel}$ corresponds to the energy caused by internal drive to keep the continuity of velocity and $E_i^{env}$ is the energy of response to environment.
%The optimization of $E_i^{vc}$ helps the agent move with current velocity,
%and the optimization of the response to environment term $E_i^{env}$ drives the agent reacting to external stimuli from the environment.
%$w_{mk} \in \mathbb{R}$ and $w_{env} \in \mathbb{R}$ are the weights of the two energy terms.

\subsubsection{Internal Drive}
%In our model, we regard that the internal drive of each agent contributes to the continuity of velocity, and it means that the agent tends keep its velocity unchanged.
%we use a velocity continuity term to represent the internal drive for an agent to keep moving with its current velocity.
%the energy of internal drive quantifies the difference between a chosen velocity $\mathbf{v}$ and the focal agent $i$'s current velocity $\mathbf{v}_{i, t}$, and it includes both the continuity of moving direction $E_{i, dir}$ and the continuity of magnitude of velocity $E_{i, mag}$, that is
We assume that the internal drive ensures the motion smoothness, which is formulated considering the first-order and second-order derivatives of motions:
\begin{equation}
\label{equ:internal drive}
E_{i, vel} = E_{i, dir} + E_{i, acc},
\end{equation}
where $E_{i, dir} = w_{dir} \cdot e^{\rho \| \hat{\mathbf{v}} - \hat{\mathbf{v}}_{i, t} \|_2}$ is the directional continuity energy, and $w_{dir} \geq 0$ is the weight.
%We also consider the continuity of its acceleration in both direction and magnitude:
We also consider the continuity of its acceleration, 
\begin{equation}
\label{equ:acc}
    E_{i,acc} = w_{acc} \cdot \left( E_{i, adir} + E_{i, amag} \right),
\end{equation}
where $w_{acc} \geq 0$ is the weight. We use $\mathbf{a}' = \mathbf{v} - \mathbf{v}_{i, t}$ to compute the predicted acceleration so that if a $d_v$ is selected, and $\mathbf{a}$ is the corresponding acceleration in $d_v$. In Equation \ref{equ:acc}, $E_{i, adir} = e^{\rho \| \hat{\mathbf{a}}' - \hat{\mathbf{a}} \|_2}$ is designed for the continuity of the direction of acceleration, and $E_{i, amag} = \left| \| \mathbf{a}' \| - \| \mathbf{a}\| \right|$ is designed for the continuity of the magnitude. Overall, $E_{i, dir}$ minimizes the direction changes and $E_{i, acc}$ minimizes the acceleration changes compared to the real data. The regularization on both motion derivatives leads to smooth motions.
%and $E_{i, mag} = \| \|\mathbf{v}\| - \|\mathbf{v}_{i, t}\| \|_2$ is the magnitude continuity energy,
%$w_{dir}$ and $w_{mag}$ are the weights of the two energy terms. 
\subsubsection{Response to Environment}
Insects can react quickly to external stimuli, e.g. following a target, or escaping from an approaching predator. To model one stimulus, the energy of response to it is defined as:
\begin{equation}
\label{equ:response environment1}
E_{i, env} = w_{env} \cdot e^{\rho \Psi},
\end{equation}
where $w_{env} \geq 0$ is the weight, and the function $\Psi$ is defined as:
\begin{equation}
\label{equ:response psi}
\centering
\Psi = 
\left\{
\begin{array}{lr}
%e^{\rho \left( \frac{\|\mathbf{p}^\prime_{i, t+1} - \mathbf{p}_{sti, t} \|_2}{d_{goal}} - 1 \right)}, \\
\begin{split}
&\| \hat{\mathbf{v}} - \hat{\mathbf{v}}_{i, sti} \|_2, &attracted, \\
&1 - \frac{\| \mathbf{p}^\prime_{i, t+1} - \mathbf{p}_{sti, t} \|_2}{d_{danger}}, &startled,\\
\end{split}
\end{array}
\right.
\end{equation}
where $\mathbf{p}_{sti, t}$ is the stimulus' position, $\hat{\mathbf{v}}_{i, sti}$ is the direction pointing from the agent's current position to the stimulus. In addition, as Figure \ref{fig:framework} shows, an agent is startled if a potential danger is within a range defined by $d_{danger}$.

\subsection{Noise}
As the movements of swarms show strong randomness, we introduce a noise-induced control term to generate plausible trajectories.
We introduce the curl noise \cite{bridson2007curl}, which has been used in force-based models to generate more accurate simulation results \cite{wang2015bswarm, ren2016simulating}, to model the motion randomness. The function $f_{N}$ denotes the curl noise function for generating a noise direction $\hat{\mathbf{v}}^n_{i, t}$. Similar to the calculation of the energy of direction continuity in Equation \ref{equ:internal drive}, the energy of noise-induced direction control is defined as:
\begin{equation}
\label{equ:nosie}
E_{i, n} = w_{n} \cdot e^{\rho \| \hat{\mathbf{v}} - \hat{\mathbf{v}}^n_{i, t} \|_2},
\end{equation}
where $w_{n} \geq 0$ is the weight of the energy term. 
%\begin{equation}
%\label{equ:noise vector}
%\begin{array}{l}
%	\mathbf{v}_{i, t}^* = f_{N}(\mathbf{s}_{i, t-1}), \\
%	\hat{\mathbf{v}}^n_{i, t} = \hat{\mathbf{v}}_{i, t}^*, \\
%	||\mathbf{v}^n_{i, t}|| = ||\mathbf{v}_{i, t} + \mathbf{v}_{i, t}^*||
%\end{array}
%\end{equation}

%Similar to the calculation of the energy of internal drive, the agents tends to move with the noise-introduced velocity $\mathbf{v}^n_{i, t}$ in Equation \ref{equ:noise vector}, and
%the energy of noise considers both the energy of noised-introduced direction $E_{i, ndir}$ and the energy of noised-introduced
%magnitude of velocity $E_{i, nmag}$, that is
%\begin{equation}\label{formula:noise energy}
%E_{i, n} = E_{i, ndir} + E_{i, nmag},
%%E_i^{n} = w_n^d e^{\rho \| \hat{\mathbf{v}}^n_{i, t} - \hat{\mathbf{v}} \|_2} + w_n^m \| \|\mathbf{v}\| - \|\mathbf{v}^n_{i, t}\| \|_2,
%\end{equation}
%where $E_{i, ndir} = e^{\rho \| \hat{\mathbf{v}}^n_{i, t} - \hat{\mathbf{v}} \|_2}$ and
%$E_{i, nmag} = \| \|\mathbf{v}\| - \|\mathbf{v}^n_{i, t}\| \|_2$.
%The optimization of the noise energy helps the agent choosing a velocity with a random change of both the direction and the speed.
\subsection{User-Defined Motion Control}
Besides the behavioral modeling of swarms, it is necessary to induce user-control for the purpose of animation. We model user-control as a direction control signal and introduce an energy term to constrain an agent to follow predefined user-defined trajectories. For agent $i$, the user-control energy is calculated as:
\begin{equation}
\label{equ:direction control}
E_{i, user} = w_{user} \cdot e^{\rho ||\hat{\mathbf{v}} - \hat{\mathbf{v}}^{cd}_{i, t}||_2},
\end{equation}
where $w_{user} \geq 0$ is the weight and $\hat{\mathbf{v}}^{cd}_{i, t}$ is the control direction of the agent for generating specific trajectories.
%this energy term calculates the difference between a chosen velocity's direction vector $\hat{\mathbf{v}}$ to the agent's current control direction $\hat{\mathbf{v}}^{cd}_{i, t}$.
\section{Results and Evaluations}
\label{sec:results and evaluations}

The implementation is done in C++ and the experiments were run on a PC with an Intel (R) Core (TM) i7 4.00 GHz CPU, 32 GB RAM, and an NVIDIA Geforce GTX 1060 GPU.
We provide both qualitative and quantitative evaluations to demonstrate the performance of FASTSWARM. Due to the space limit, we only show representative results and refer the readers to the supplementary materials for more details. In all our experiments, we set $\rho = 2.5$, and the weights of the energy terms for the test scenarios are shown in Table \ref{tab:weights}.
%and Table \ref{tab:example time} shows their time performances. {\color{red} You give performance results in two parts. Perhaps combine them into one sub section?}

\subsection{Qualitative Results}
\label{sec:subsec:results}
We first show qualitative results. The results are divided into two parts: natural behaviors and user-controlled behaviors, the former showing the high visual realism and the latter showing controllability. 

\subsubsection{Collective Behaviors}
\label{sec:subsec:collective behaviors}
We use the reference dataset from \cite{kelley2013emergent} to generate the \textit{aggregation}, \textit{mating}, and \textit{escaping} behaviors of flying insect swarms. The results show that FASTSWARM is capable of simulating a variety of natural behaviors with good visual quality.

\begin{figure}[t]
	\centering
	\subfigure[37 insects]{
		\label{subfig:aggregation50} 
		\includegraphics[width=0.45\linewidth]{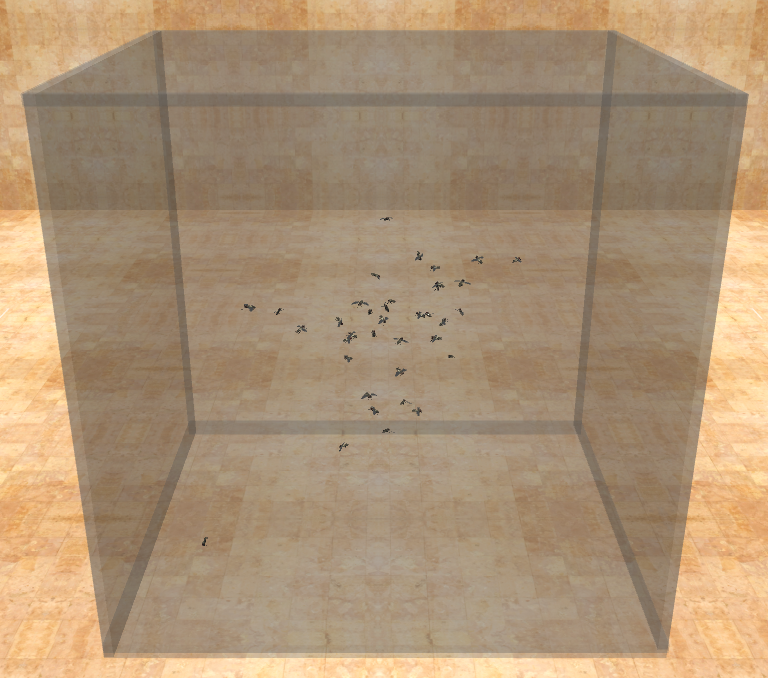}}
	\hspace{0.05in}
	\subfigure[300 insects]{
		\label{subfig:aggregation300} 
		\includegraphics[width=0.45\linewidth]{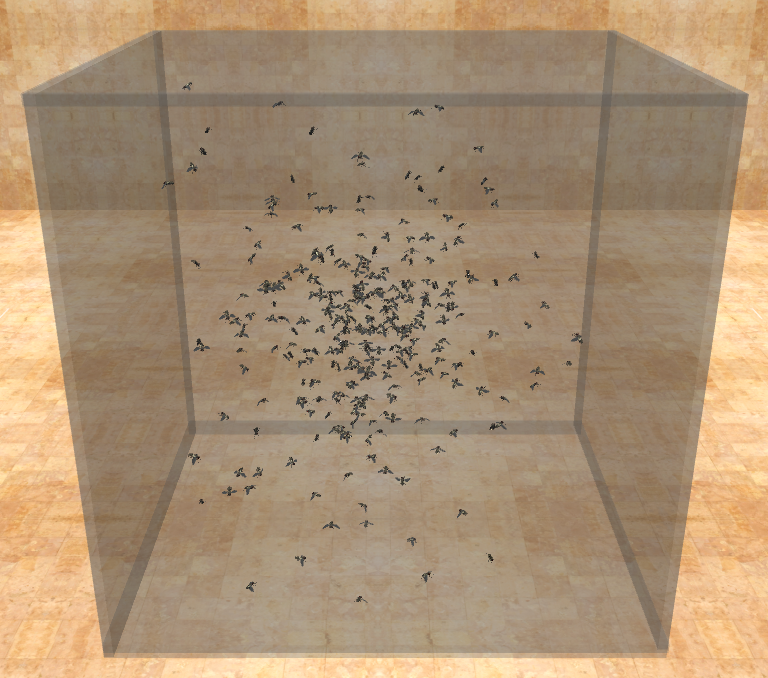}}
	\caption{\label{fig:aggregation} \textit{Aggregation} behavior of insect swarms in different scales.
% 	The flying insects in Figure \ref{subfig:aggregation50} or Figure \ref{subfig:aggregation300} are attracted by its high-density swarm center, and the swarm center almost remains unchanged.
	}
\end{figure}

\textbf{Aggregation}. Our approach can generate the \textit{aggregation} behavior of insect swarms in different scales (see Figure \ref{fig:aggregation}). Figure \ref{subfig:aggregation50} and Figure \ref{subfig:aggregation300} respectively shows two swarms of different sizes in the aggregation scenario of the reference dataset with an unchanged swarm center.
%During simulation, the agents move plausibly but aggregate around the swarm center. {\color{red} the last sentence is dangerous. How can you tell it is plausible?}

\begin{figure}[h]
\centering
\begin{minipage}[t]{0.45\linewidth}
\centering
\includegraphics[width=\linewidth]{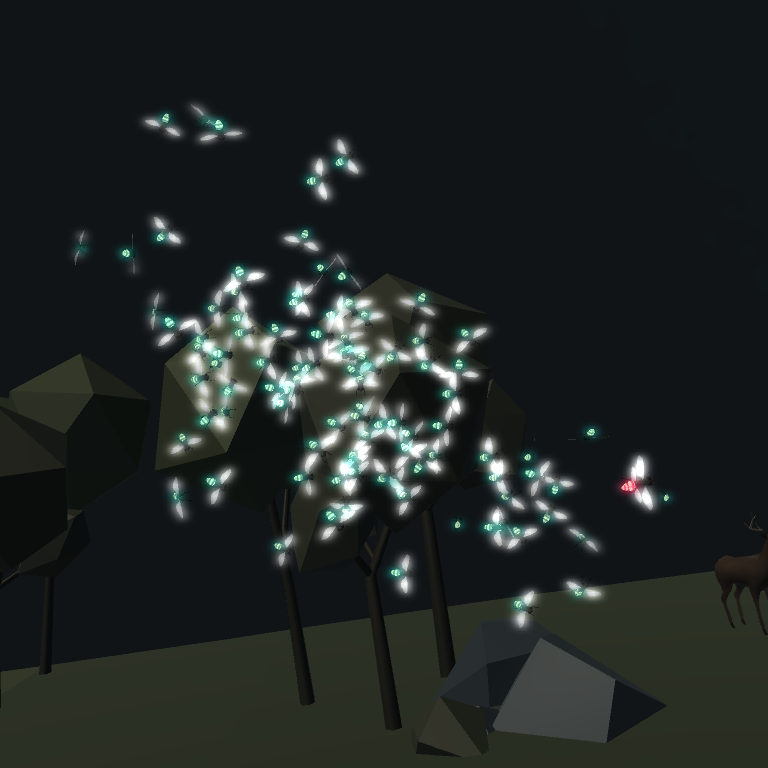}
\caption{\label{fig:mating_behavior}
\textit{Mating} behavior.}
\end{minipage}
\hspace{0.05in}
\begin{minipage}[t]{0.45\linewidth}
\centering
\includegraphics[width=\linewidth]{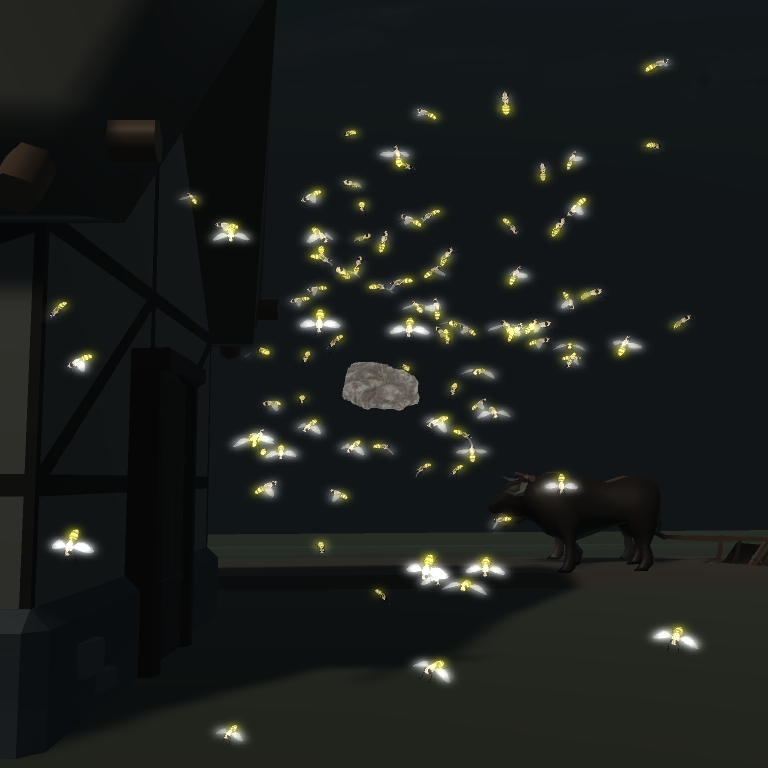}
\caption{\label{fig:escaping_behavior}
\textit{Escaping} behavior.}
\end{minipage}
\end{figure}

\textbf{Mating}. Figure \ref{fig:mating_behavior} shows the \textit{mating} behavior of insects generated by our approach. 100 male agents (the cyan ones) are attracted by a female (the red one) insect and run after the female.
% During simulation, male agents move plausibly and follow the female agent.
% In this example, the female agent is initialized with a random motion state, and each male one is initialized with a random position near the female.

\textbf{Escaping}. Our approach can also generate the \textit{escaping} behavior of insect swarms.
In Figure \ref{fig:escaping_behavior}, 100 flying insects are startled by a sudden obstacle, and try to escape from the obstacle.
%In this example, we initialize the swarm center with a random position, and initialize each agent in the swarm with a random position near the swarm center. We also set the external stimulus with a random motion state. 
During simulation, the obstacle will pass through the swarm, and the insects in the swarm escape from the danger and then aggregate again.

\subsubsection{User-Defined Shape Constraint}
\begin{figure}[h]
	\centering
	\subfigure[500 flying insects form a heart shape.]{
		\label{fig:subfig:heart} 
		\includegraphics[width=0.45\linewidth]{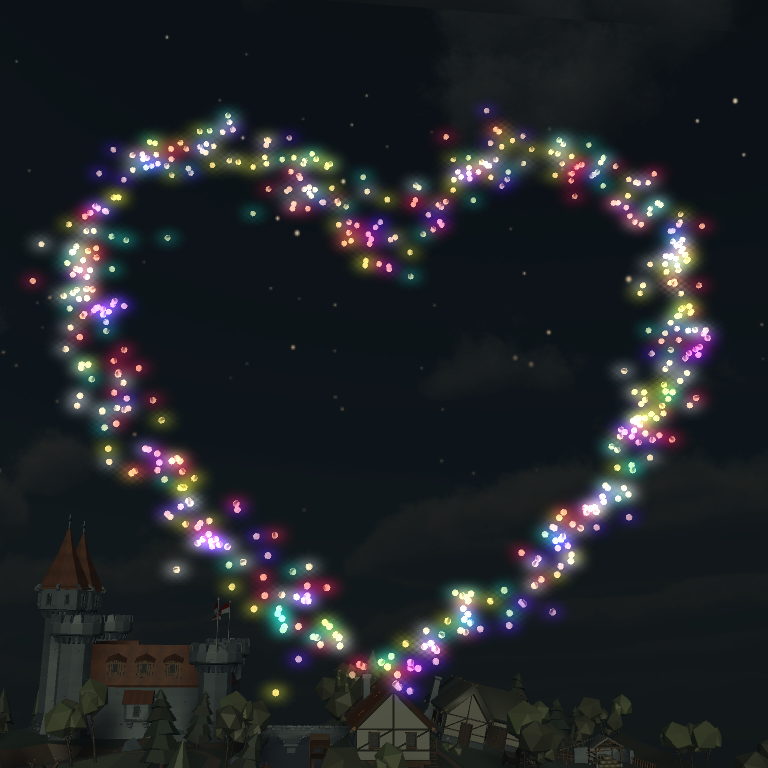}
	}
\hspace{0.05in}
	\subfigure[500 flying insects form a star shape.]{
		\label{fig:subfig:star} 
		\includegraphics[width=0.45\linewidth]{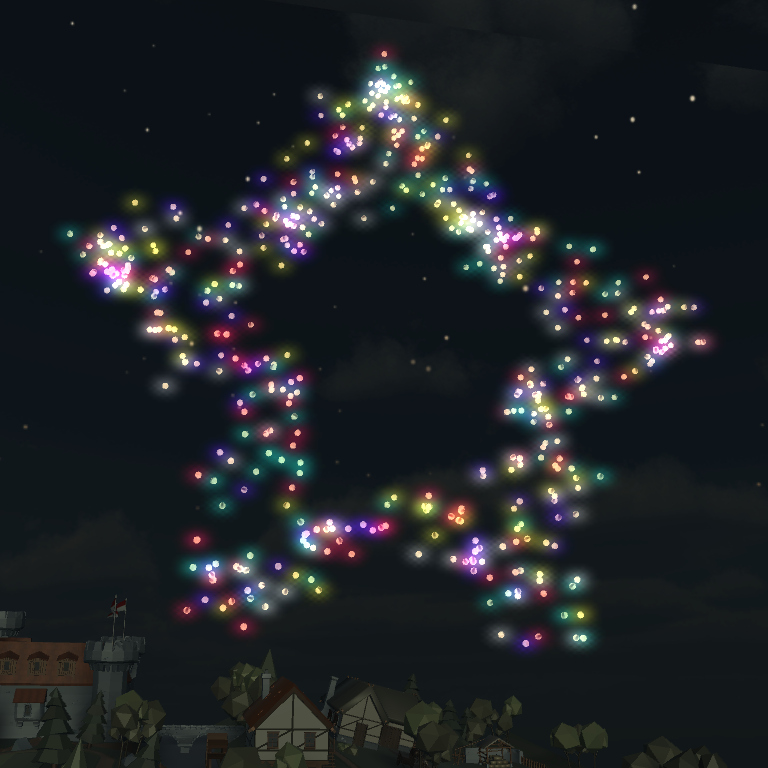}
	}
	\caption{\label{fig:shape_constraint}Two shape-constrained flocking examples. There are 11 key points in (a) and 10 key points in (b).}
\end{figure}

To evaluate the controllability, we ask the users to draw different 3D shapes which are used as constraints. Figure \ref{fig:shape_constraint} shows two shape-constrained examples to constrain the agents to move following different 3D curves. We regard the 3D curve as a bidirectional curve with several keypoints (Figure \ref{fig:framework}: User-Defined Control). Each agent is randomly initialized near a key point with a random velocity selected from the referenced dataset. Then, it is assigned a task to traverse the keypoints. During simulation, we update the control direction $\hat{\mathbf{v}}_{i, t}^{cd}$ of agent $i$ as the direction from the agent's position $\mathbf{p}_{i, t}$ to the position of the goal keypoint $\mathbf{p}_{i, t}^{goal}$:
\begin{equation}
\label{equ:control direction}
\begin{array}{l}
\mathbf{V}_{i, goal} = \mathbf{p}_{i, t}^{goal} - \mathbf{p}_{i, t}, \\
\hat{\mathbf{v}}_{i, t}^{cd} = \hat{\mathbf{V}}_{i, goal},
\end{array}
\end{equation}
where $\hat{\mathbf{V}}_{i, goal}$ is the directional vector of $\mathbf{V}_{i, goal}$. The goal keypoint will be updated to the next keypoint when the agent reaches its current goal.
\begin{table*}[t]
\centering
\small
\begin{tabular}{|c|c|c|c|c|c|c|c|c|c|}
\hline
\multicolumn{2}{|c|}{Scenario}                                                                  & $w_{rep}$          & $w_{oa}$ & $w_{cos}$ & $w_{dir}$          & $w_{acc}$          & $w_{env}$           &  $w_{n}$       & $w_{user}$         \\ \hline \hline
\multicolumn{2}{|c|}{\begin{tabular}[c]{@{}c@{}}Aggregation\\ ($N$ = 37 / 300)\end{tabular}}      & 1                  & 1        & 0.05      & 1                  & 1                  & 0                   & 0.2                  & 0                  \\ \hline
\multicolumn{2}{|c|}{Mating}                                                                    & 1                  & 1        & 0.05      & 1                  & 1                  & 1                   & 0.5                  & 0                  \\ \hline
\multirow{2}{*}{Escaping}                              & Aggregation                            & \multirow{2}{*}{1} & 1        & 0.2       & \multirow{2}{*}{1} & \multirow{2}{*}{1} & \multirow{2}{*}{2} & \multirow{2}{*}{0.3} & \multirow{2}{*}{0} \\ \cline{2-2} \cline{4-5}
                                                       & In danger                              &                    & 0        & 0      &                    &                    &                     &                      &                    \\ \hline
\multicolumn{2}{|c|}{\begin{tabular}[c]{@{}c@{}}Shape Constraint\\ (Heart / Star)\end{tabular}} & 1                  & 0.5        & 0         & 1                  & 1                  & 0                   & 1                  & 1                  \\ \hline
\end{tabular}
\caption{The weights for our implementation in Section \ref{sec:subsec:results}.}
\label{tab:weights}

\end{table*}

\begin{table}[t]
	\centering
	\small 
\begin{tabular}{|c|c|l|c|}
\hline 
Scenario                                                                  & \multicolumn{2}{c|}{$N$} & Time (s/f) \\ \hline \hline
\multirow{2}{*}{Aggregation}                                              & \multicolumn{2}{c|}{37}  & 0.0003     \\ \cline{2-4} 
                                                                          & \multicolumn{2}{c|}{300} & 0.0025     \\ \hline
Mating                                                                    & \multicolumn{2}{c|}{100} & 0.0159     \\ \hline
Escaping                                                                  & \multicolumn{2}{c|}{100} & 0.0107     \\ \hline
\begin{tabular}[c]{@{}c@{}}Shape Constraint\\ (Heart / Star)\end{tabular} & \multicolumn{2}{c|}{500} & 0.0037     \\ \hline
\end{tabular}
\caption{\label{tab:example time}
	Time performance of the simulation results shown in Section \ref{sec:subsec:results}}
\end{table}
\subsection{Quantitative Evaluations}
\subsubsection{Time Performance}
\begin{figure}[htb]
	\centering
	\includegraphics[width=0.8\linewidth]{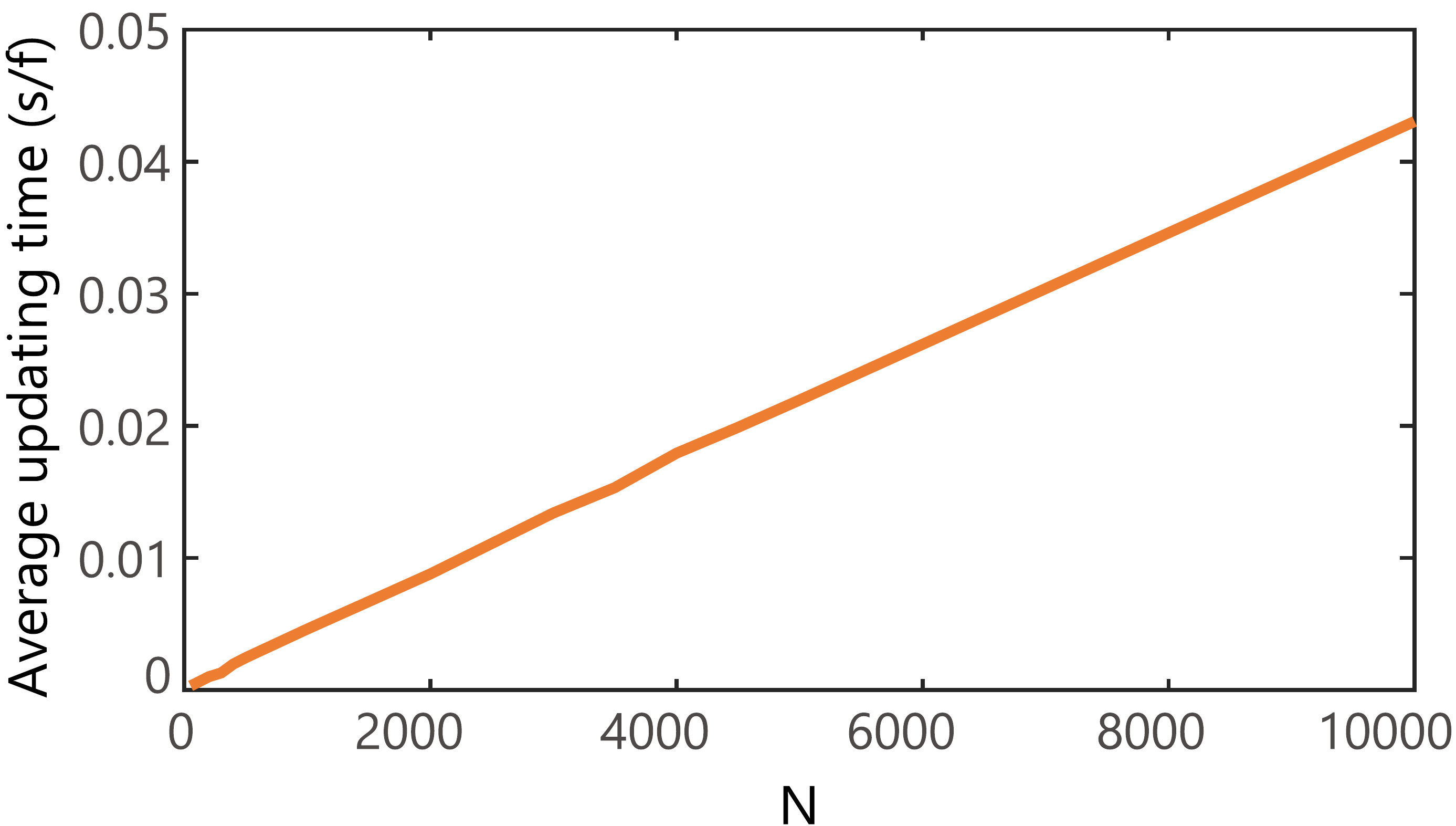}
	\caption{\label{fig:timePerformance}
		Time performance of our approach. The computation cost is linear w.r.t the number of insects}
\end{figure}

To quantitatively test the scalability, we evaluate FASTSWARM with an increasing number of insects in an \textit{aggregation} scenario, similar to Section \ref{sec:subsec:collective behaviors}. Table \ref{tab:example time} shows the time performance of the test scenarios in Section \ref{sec:subsec:results}. Figure \ref{fig:timePerformance} shows the time performance of our method against different swarm scales.
% In our implementation, we accelerate our approach using the following three ways.
Theoretically, the time complexity is $O(kN)$ where $N$ is the number of insects and $k$ is data sample size in the reference data. In our experiments, $k$ = 300 is enough to generate all behaviors. The linear time complexity guarantees the high-performance of FASTSWARM. In practice, we further accelerate the computation in three ways:
\begin{itemize}
\item Fast indexing. We discretize the space into a 3D grid and index the neighborhood region of every agent, and approximate the repulsion and attraction regions using cubes whose dimension is the distance thresholds in their respective energy terms. During simulation, only local search and computation is done. 
% accelerate the spatial neighborhood search using a 3D grid data structure. We organize the whole motion scene as a 3D grid, and each cell in the grid has its own index based on its covered 3D region. At each timestep, an agent only belongs to one cell based on its current location.
% When calculating the repulsion and attraction energy terms, we reduce the cost of neighborhood search by taking the spatial region of repulsion or attraction as a cube whose edge length is the distance threshold. In this way, we can only search agents locating in the sub cells within the \textit{repulsion zone} or \textit{attraction-agent zone}.

\item Reduced search space. We sort the reference dataset by the speed and divide the dataset into several groups. To choose a new velocity for an agent, our algorithm only traverses the corresponding group and the groups with similar speed.

\item Parallelization. As Equation \ref{equ:updating function} is computed for each agent and is therefore highly parallelizable. We also parallelize FASTSWARM to concurrently compute the updates for multiple agents.
\end{itemize}

\subsubsection{Comparisons}
\begin{figure*}[t]
 	\centering
 	\includegraphics[width=\linewidth]{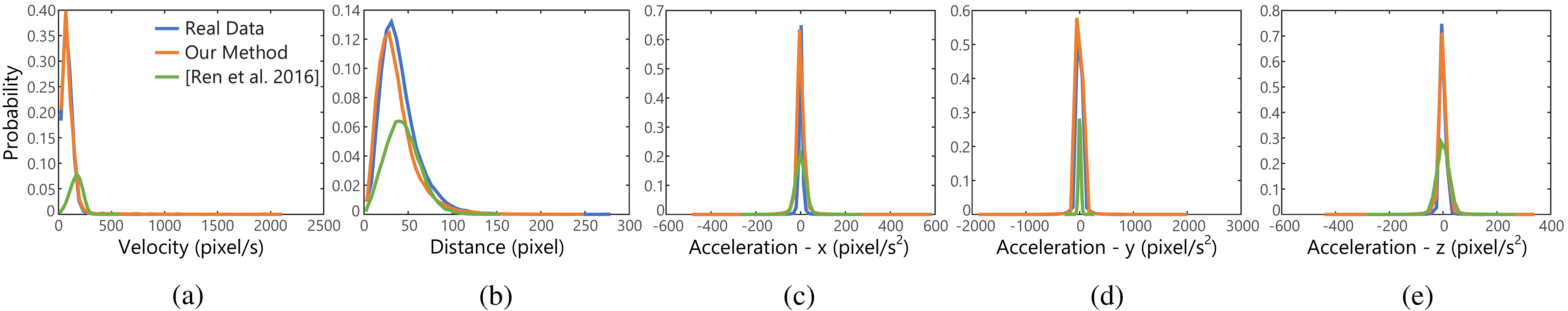}
 	\caption{\label{fig:comparison}
 	Comparisons of the distributions of the velocity (a), minimum distance (b), and acceleration (c-e for $x$, $y$ and $z$ direction) between our method, the real data, and \cite{wang2015bswarm}.
 	}
\end{figure*}

We compare our method with the method in \cite{wang2015bswarm} and the ground-truth under different metrics. Here, we take the real tracklets \cite{wu2011automated} as the ground-truth. 
In both comparisons, the scenes are the same aggregation scenario in a box as the ground-truth \cite{wu2011automated}, and the swarms are initialized by randomly selecting one frame from the real data.
% The initialization processes of both comparison experiments are the same {\color{red} where the initial motion state of the swarm is randomly selected from the real data}. In addition, the motion scenes are the same as the ones in the ground-truth {\color{red} where the insects are aggregating in a cubic box}. 
Since directly comparing individual trajectories is not possible, the evaluation metrics include distributions of density, velocity and acceleration, as they capture both the state and the motion dynamics. Similar to previous work \cite{ren2019heter}, we use minimum-distance to describe the density. Results are shown in Figure \ref{fig:comparison}. The distributions of velocity, minimal distance and acceleration of FASTSWARM are much closer to those in the ground-truth than \cite{wang2015bswarm}. This is true for both motion dynamics (velocity in Figure \ref{fig:comparison}(a) and acceleration in Figure \ref{fig:comparison}(c-d)) and states (density in Figure \ref{fig:comparison}(b)). The better motion dynamics shows that FASTSWARM is superior in capturing realistic behaviors.
% \textbf{Velocity}. As velocity is a basic feature of motion dynamics, we report the comparisons the distributions of velocities in Figure \ref{fig:comparison}(a).
% \textbf{Density}. Density distribution is a key feature for swarm behaviors. As minimum distance (the distance to the nearest agent) is useful to describe the density, we report the comparisons of the distributions of minimum distances in Figure \ref{fig:comparison}(b).
% \textbf{Acceleration}. Since acceleration can be considered as the effective net force on an agent for decision making of the movement, we report the comparisons of the distributions of acceleration along three axises in Figures \ref{fig:comparison}(c-e).
% The closer distributions of velocity and acceleration shown in Figure \ref{fig:comparison}(a, c-d) demonstrate that the motion update performed by Equation \ref{equ:updating function} works well than the prior work \cite{wang2015bswarm} on capturing the motion characteristic of an insect swarm.

For comparing the density, in \cite{wang2015bswarm}, the density of the swarm is controlled simply by a distance-based attraction force, and the homogeneous distance control results in the relatively uniform density of the swarm as the agents aggregate around the attraction boundary. Addressing this issue, we control the aggregation of agents by combining the distance-based attraction among agents with a direction control that drives the agents to move towards the swarm center, and Figure \ref{fig:comparison}(b) shows that our method performs better in controlling the density of the swarm to be similar to the reference dataset.

\subsubsection{Trajectory Synthesis}
\begin{figure}[h]
	\centering
	\includegraphics[width=0.75\linewidth]{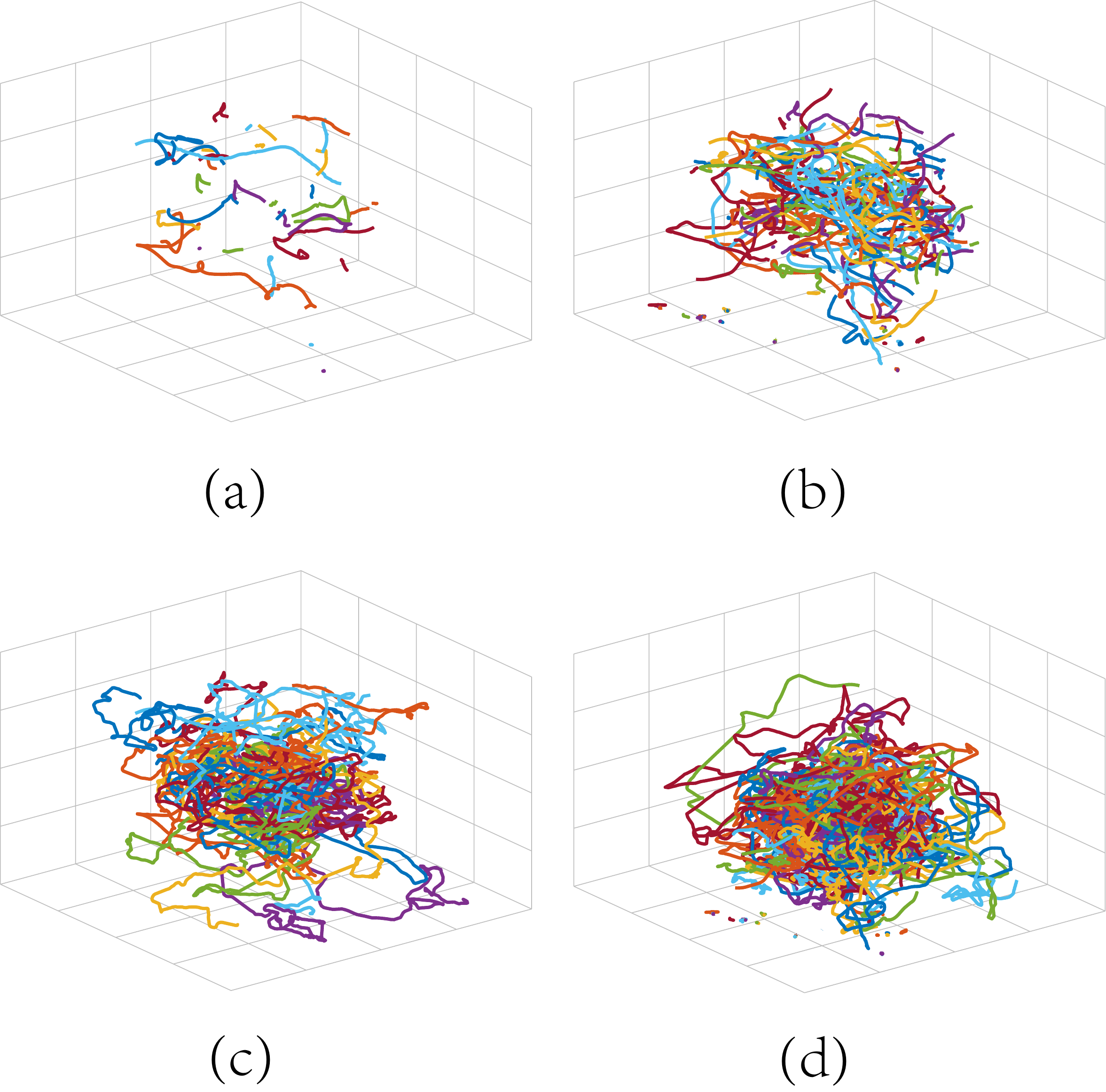}
	\caption{\label{fig:traj_completion} Results of trajectory synthesis. (a) shows the captured intermittent trajectories of 37 agents. (b) shows the real trajectories of the following 500 frames where exist lots of intermittent segments. (c) shows our prediction result of (a) by extending the initial trajectories to 500 frames. (d) shows the blending trajectories of 100 virtual agents with (b).}
\end{figure}

% {\color{red} Based on the real data, FASTSWARM can be used to recover tracklets into trajectoies or augment the dataset to achieve global visual similarity. However, the numerical accuracy is difficult to compute as there are only tracklets in the original data.}
% As captured trajectories of insect swarms usually contain missing data because insects tracking is rather difficult, our approach provides a new means to predict the trajectories of incomplete trajectory segments.
Given only trackelets (instead of whole trajectories) are avaiable, FASTSWARM provides a possible avenue to recover them for animation applications. Although it is difficult to exactly recover the trajectories or even compute their numerical accuracy, due to the lack of ground-truth data, our recovered trajectories can achieve global visual similarities. As shown in Figure \ref{fig:traj_completion}, we extract the trajectory segments of 37 agents from the real-world dataset of \cite{kelley2013emergent} (Figure \ref{fig:traj_completion}(a)). By taking these trajectory segments as the initial trajectories of a swarm with 37 agents, our method can be used to predict the subsequent trajectories (Figure \ref{fig:traj_completion}(c)). 

Our method can also synthesize plausible insect swarm animations by mixing the trajectories from the real-world dataset with the synthetic trajectories. As Figure \ref{fig:traj_completion}(d) shows, in the real-world trajectories in \ref{fig:traj_completion}(b), we add 100 agents that are simulated by FASTSWARM.
% The mixed trajectories are visually similar to the original data. {\color{red}I am a bit concerned here as there is no numerical comparison between your synthesised trajectories with real trajectories. If you claim you can fix the missing trajectories, you will need to do numerical comparisons.}
\section{Conclusion}
\label{sec:conclusion}
We have presented a general and scalable data-driven optimization framework to simulate flying insect swarms in real time. Our optimization method is capable of generating natural collective behaviors of flying insects by utilizing a motion characteristic dataset extracted from real data. The generated animations are plausible and have high visual realism. We have validated our approach using extensive experiments through qualitative and quantitative analysis. Our method also provides a means to generate interesting shape-constrained swarm behaviors controlled by users, which benefits computer animation. Moreover, our method can be used to predict the missing trajectories in the captured dataset and augment the dataset with different number of insects while maintaining the global visual similarity to the real data.
% To demonstrate the plausibility of our simulation results, we compare our method with the prior work \cite{wang2015bswarm} in the same scenarios with the real-world dataset, and the quantification comparisons show that our result is more closer to the real-world dataset (Figure \ref{fig:comparison}). Our method can generate diverse simulation results that are different from the original real-world dataset. For example, we can generate different scales of agents in the same environment as the original dataset (Figure \ref{fig:aggregation}), we can generate various behaviors of flying insect swarms by referencing a simple dataset (Figure \ref{fig:mating_behavior} and Figure \ref{fig:escaping_behavior}), we can also enforce user-defined virtual behaviors of agents (Figure \ref{fig:shape_constraint}). Our method can also be used as a tool to predict trajectories for the fragmentary segments of flying insect swarms or blend virtual agents with real data (Figure \ref{fig:traj_completion}). In addition, our method can be used for interactive simulations for different scales of agents in different scenarios (Table \ref{tab:example time}).

\textbf{Limitations and Future Work}. Our method relies on the quality of the reference dataset. However, as our reference dataset is extracted from imperfect trajectories captured using computer vision techniques, which contain noises and erroneous data, our result may replicate the defects in the input data. In the future, we are interested in exploring a more general framework so that it can simulate any types of motion (not restricted to insects) with limited captured trajectories from real world or created by animators. We are also planning to employ the Long Short-Term Memory (LSTM) networks to explore the latent representations of trajectories to further enhance our results. 

%\input{acknowledgement.tex}
%-------------------------------------------------------------------------

\begin{small}
\bibliographystyle{unsrt}
\bibliography{reference}
\end{small}

\end{document}